\documentclass{emulateapj}
\slugcomment{{\sc Published in ApJL:} July 10, 2011} 

\usepackage{ulem}
\usepackage{color}
\usepackage{graphicx}
\usepackage{subfigure}
\usepackage{footnote}
\makesavenoteenv{tabular}

\shorttitle{Spatially-Resolved Dual AGN}
\shortauthors{McGurk et al.}

\newcommand{\obj}{J0952+2552}
\newcommand{\kms}{km s$^{-1}$}
\newcommand{\hb}{H$\beta$}
\newcommand{\ha}{H$\alpha$}
\newcommand{\nii}{[\ion{N}{2}]}
\newcommand{\pab}{Pa$\beta$}
\newcommand{\siii}{[\ion{S}{3}]}
\newcommand{\feii}{[\ion{Fe}{2}]}
\newcommand{\feiij}{[\ion{Fe}{2}] $\lambda 9188$}
\newcommand{\feiih}{[\ion{Fe}{2}] $\lambda 1.2570\mu$m}
\newcommand{\oi}{\ion{O}{1}}
\newcommand{\siiis}{[\ion{S}{3}] $\lambda 9534$}
\newcommand{\siiiw}{[\ion{S}{3}] $\lambda 9073$}
\newcommand{\oiiis}{[\ion{O}{3}] $\lambda 5008$}
\newcommand{\oiiiw}{[\ion{O}{3}] $\lambda 4959$}
\newcommand{\oiii}{[\ion{O}{3}]}

\begin{document}

\title{Spatially Resolved Spectroscopy of SDSS J0952+2552: a confirmed Dual Active Galactic Nuclei }

\author{ R.C. McGurk\altaffilmark{1}, C.E. Max\altaffilmark{1}, D.J. Rosario\altaffilmark{2}, 
G.A. Shields\altaffilmark{3}, K.L. Smith\altaffilmark{3}, and S.A. Wright\altaffilmark{4}}

\altaffiltext{1}{Astronomy Department and UCO-Lick Observatory, University of California, Santa Cruz, CA 95064, USA; rmcgurk@ucsc.edu, max@ucolick.org}
\altaffiltext{2}{Max-Planck-Institute for Extraterrestrial Physics, Garching, 85748, Germany; rosario@mpe.mpg.de}
\altaffiltext{3}{Astronomy Department, University of Texas, Austin, TX 78712, USA; shieldsga@mail.utexas.edu, krista@mail.utexas.edu}
\altaffiltext{4}{Astronomy Department, University of California, Berkeley, CA 94720, USA; saw@astro.berkeley.edu}

\begin{abstract}

Most massive galaxies contain supermassive black holes (SMBHs) in their cores. When galaxies merge, gas is driven to nuclear regions and can accrete onto the central black hole. Thus, one expects to see dual active galactic nuclei (AGNs) in a fraction of galaxy mergers. Candidates for galaxies containing dual AGNs have been identified by the presence of double-peaked narrow [O III] emission lines and by high spatial resolution images of close galaxy pairs.  Spatially resolved spectroscopy is needed to confirm these galaxy pairs as systems with spatially separated double SMBHs.  With the Keck 2 Laser Guide Star Adaptive Optics system and the OH Suppressing InfraRed Imaging Spectrograph near-infrared integral field spectrograph, we obtained spatially resolved spectra for SDSS J09527.62+255257.2, a radio-quiet quasar shown by previous imaging to consist of a galaxy and its close (1\farcs0) companion. We find that the main galaxy is a Type 1 AGN with both broad and narrow AGN emission lines in its spectrum, while the companion galaxy is a Type 2 AGN with narrow emission lines only. The two AGNs are separated by 4.8 kpc, and their redshifts correspond to those of the double peaks of the [O III] emission line seen in the Sloan Digital Sky Survey spectrum.  Line diagnostics indicate that both components of the double emission lines are due to AGN photoionization.  These results confirm that J0952+2552 contains two spatially separated AGNs.  As one of the few confirmed dual AGNs at an intermediate separation of \textless 10 kpc, this system offers a unique opportunity to study galaxy mergers and their effect on black hole growth.

\end{abstract}

{\it Facilities:} \facility{Keck:II (laser guide star adaptive optics, OSIRIS)}

\section{Introduction} 

In $\Lambda$CDM cosmology, galaxies regularly interact and merge; hierarchical galaxy formation and evolution dictates that galaxies undergo multiple mergers as they evolve \citep[e.g.,][]{blumenthal1984, spergel2007}.  We now know that most massive galaxies harbor supermassive black holes (SMBHs) in their centers \citep{richstone1998, kormendy1995}. One picture explaining active galactic nuclei (AGNs) invokes mergers of gas-rich galaxies; the gravitational perturbations due to the merger allow gas to flow into the galactic center and trigger accretion onto the black hole \citep{hernquist1989, kauffmann2000, hopkins2008}.  A complication in this picture is the AGN's duty cycle, or how long each AGN remains active during the merger.  If AGN activity is triggered by galaxy mergers, AGN pairs should be observable in at least some galaxy mergers.  Thus, the existence and statistics of dual AGNs provide an important probe into hierarchical galaxy formation models, accretion-triggering mechanisms, galaxy merger rates, and SMBH growth \citep[and references therein]{yu2011arxiv}.

AGN pairs have been found and resolved both spatially and spectrally at a wide range of separations.  Hundreds of AGN pairs are known at \textgreater 10 kpc separations \citep[e.g.,][]{myers2007, myers2008, hennawi2010, green2010, piconcelli2010}, while only three candidate binary AGNs are known at \textless 10 pc separations \citep{rodriguez2006, boroson2009, decarli2010}.  Only five spatially and spectrally confirmed dual AGNs are known with intermediate separations of between 0.1 and 10 kpc: LBQS 0103-2753 \citep{junkkarinen2001}; NGC 6240 \citep{komossa2003}; Arp 299 \citep{ballo2004}; J1420+5259 \citep{gerke2007}; Mrk 463 \citep{bianchi2008}; Mrk 739 \citep{koss2011}; and J1027+1749 \citep{liu2011}. Eighteen candidate AGN pairs with spatially unresolved double-peaked emission lines and resolved double spatial structure have been observed by \citet{fu2011} and \citet{ rosario2011}.  

Potential dual AGNs have been selected using the following two methods.
\begin{enumerate}
    \item Observing velocity offsets between emission lines, such as \oiii, potentially from two AGNs \citep{smith2010, zhou2004, wang2009, liu2010a, liu2010b, comerford2009a, comerford2009b, XuKomossa2009, gerke2007}.
    \item Imaging multiple potential AGNs in or close to a single host galaxy \citep{junkkarinen2001, komossa2003, comerford2009b, fu2011, rosario2011, liu2011}.
\end{enumerate}
To \underline{confirm} a potential dual AGN, spatially resolved spectroscopy is needed to prove that each resolved source has a unique AGN spectrum.  Without spatially resolved spectroscopy to match the observed galaxies to their corresponding double-peaked emission lines, the double peaks of spectral lines may be due to a chance superposition of two objects, a recoiling SMBH \citep{bonning2007, civano2010, guedes2011},  jets interacting with the surrounding medium \citep{rosario2010}, outflows from a single AGN \citep{fischer2011}, or rings of star formation.   Similarly, without spatially resolved spectroscopy, the multiple bright cores imaged in or around one galaxy may be a chance superposition, a recoiling SMBH, jets interacting with the surrounding medium, outflows from a single AGN, gravitationally lensed sources \citep[and references therein]{hennawi2006}, or starbursts.  High-energy X-ray observations are another unambiguous way to confirm true dual AGNs.

In the next section, we discuss W. M. Keck Observatory adaptive optics imaging of \obj, a potential pair of bright AGNs with double-peaked emission lines in the Sloan Digital Sky Survey (SDSS) spectrum. In Section 3, we describe our spectroscopic observations of \obj~using adaptive optics spectroscopy from Keck's OH Suppressing InfraRed Spectrograph (OSIRIS) and the resulting data reduction.  In Section 4, we match resolved spatial components seen by NIRC2 and OSIRIS to the SDSS double-peaked emission lines, and discuss what type of physical mechanisms are generating these emission lines.  Section 5 presents our conclusions.   We adopt a concordance cosmology with $H_{0} = 70$ km s$^{-1}$ Mpc$^{-1}$ and $\Omega_{\Lambda} = 0.7$.  All quoted wavelengths are in vacuum units.

\section{ Imaging of \obj} 

In \citet{rosario2011}, we undertook a program of near-infrared (NIR) imaging of candidate double AGNs using the Keck Laser Guide Star Adaptive Optics (LGS AO) system \citep{wizinowich2006}.  For a well-defined sample of candidate dual AGNs, \citet{rosario2011} drew on the catalog of \citet{smith2010}, which selected SDSS DR7\footnote{http://www.sdss.org/dr7/} spectroscopic AGNs that show a double-peaked \oiiis~line.  From this catalog, \citet{rosario2011} selected a sample of radio-undetected, optical Type 1 AGNs observable with AO. Using the NIRC2 camera\footnote{http://www2.keck.hawaii.edu/inst/nirc2/}, \citet{rosario2011} found that out of the 12 targets imaged, 6 (50\%) were dual AGN candidates.  Four candidates were in close mergers and two were in more distant pairs with separations ranging from 3 to 12 kpc.  The recent study by \citet{fu2011} imaged a larger sample of both radio-loud and radio-quiet Type 1 and 2 AGNs to a shallower depth, and found that a similar fraction of their AGNs were spatially separated galaxy pairs.

\begin{figure}[t]
\figurenum{1}
\label{images}
\centering    
\includegraphics[width=\columnwidth]{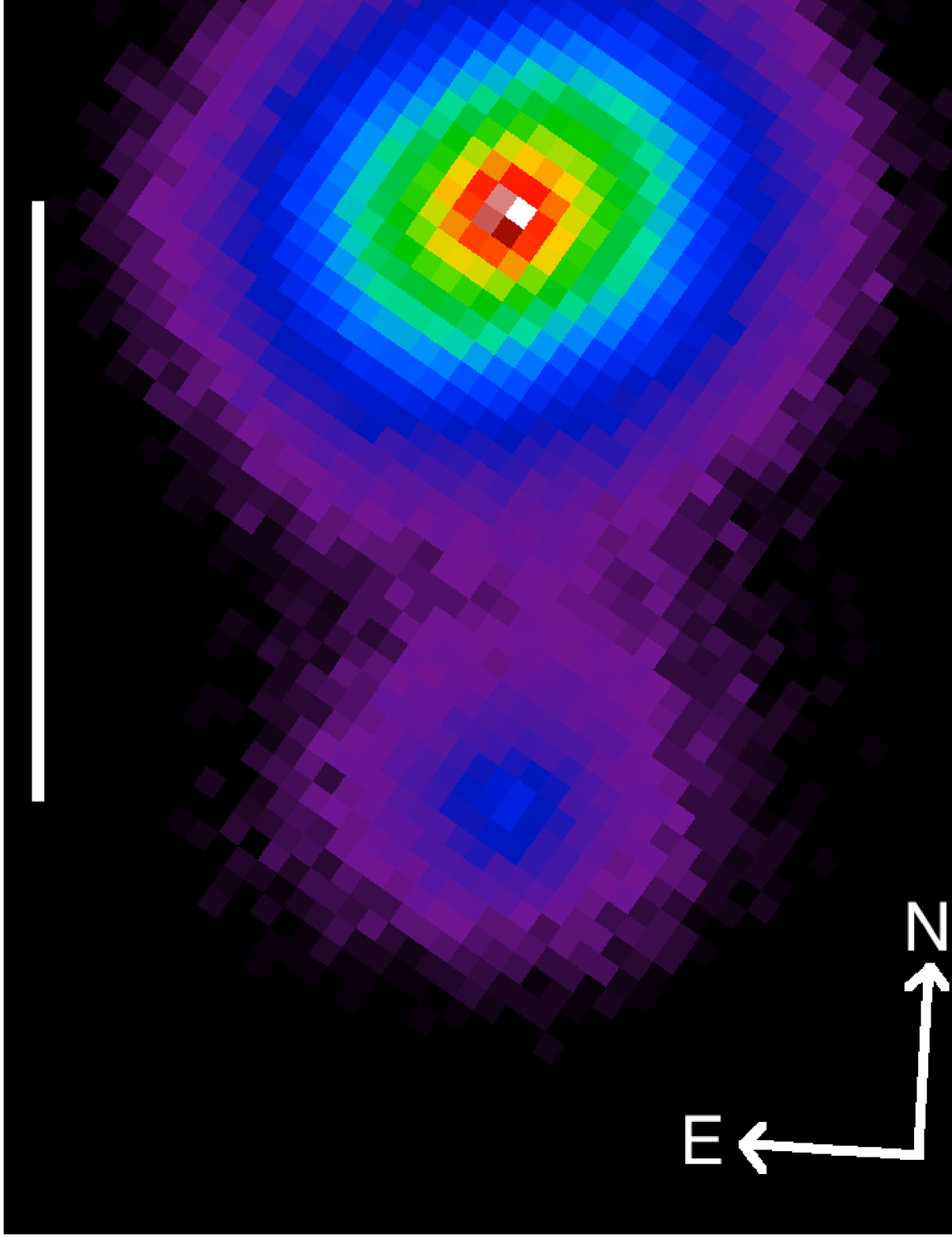}
\caption{ Keck LGS AO imaging of \obj, a Type 1 AGN with double-peaked narrow lines.  The white line in both images corresponds to 1\farcs0 or 4.80 kpc at the redshift of the system. Left: NIRC2 wide camera (0\farcs04 pixel$^{-1}$) \textsl{H}-band ($\lambda_{cen}=1.633 \mu$m) log-scaled image. Right: OSIRIS \textsl{H}-band ($\lambda_{cen}=1.638 \mu$m) log-scaled image, created by summing up the flux over all the wavelengths in each of the spatial pixels.  Each OSIRIS spatial pixel has a width of 0\farcs1 or 0.48 kpc at the redshift of the system.  The bright main galaxy and its companion are clearly visible.
}
\end{figure}

However, only spatially resolved spectroscopy can confirm these six candidates as actual dual AGNs.  Here we present spatially resolved spectroscopy of one promising dual AGN candidate, SDSS J09527.62+255257.2, or J0952+2552 for short, imaged by both \citet{rosario2011} and \citet{fu2011}. Figure \ref{images} (left panel) shows the LGS AO NIRC2 \textsl{H}-band ($1.337$-$1.929 \mu$m) image of \obj.  The average redshift of \obj~is 0.339 \citep{smith2010}, and the measured separation between the bright main galaxy and the companion is 1\farcs0, or 4.81 kpc.

We modeled the NIRC2 \textsl{H}-band image of \obj~using the galaxy structure fitting code GALFIT v3.0 \citep{peng2002}.  We modeled both the main and the companion galaxies as a combination of a point source for the AGN, and an extended stellar light distribution for the galaxy with a variable S\'{e}rsic index, half-light radius, ellipticity, and orientation.  The fit was visually examined, and the initial parameters were varied until a good fit was achieved.  Both galaxies are disk-like, with S\'{e}rsic indices of $1.5\pm1$.

\section{ Observations } 

We used the OSIRIS \citep{larkin2006} and the LGS AO system on the Keck 2 telescope to obtain spatially resolved spectroscopy of \obj.  The 0\farcs1/spatial pixel (spaxel) plate scale was used to maximize throughput.  Observing with this plate scale in the \textsl{J} ($1.180$-$1.416 \mu$m) and \textsl{H} ($1.473$-$1.803 \mu$m) broadband filters results in a spectral resolution of 3000 and a field of view of 1\farcs6 $\times$ 6\farcs4.  We observed \obj~using a half-night on 2010 December 29.  Integration times totaled 100 minutes in \textsl{J}-band and 80 minutes in \textsl{H}-band.  We also imaged the tip-tilt star for Point Spread Function calibration and an A5V star as a telluric standard.  Our tip-tilt star had an \textsl{R} magnitude of 14.6 and a separation from the target of 40\farcs9.  The predicted seeing for the night was 0\farcs8.

Accurate sky subtraction is very important due to the abundance of sky lines in the NIR.  In extent the main galaxy and its companion are about 3\arcsec~long. Since the detector is 6\farcs4 long, we positioned the galaxy pair on one half of the detector and empty sky on the other half, and then nodded the telescope so that in the next exposure the sky and object positions on the detector were flipped. 

\begin{figure*}[t]
\figurenum{2}
\label{spectra}
\centering
\includegraphics[width=\textwidth]{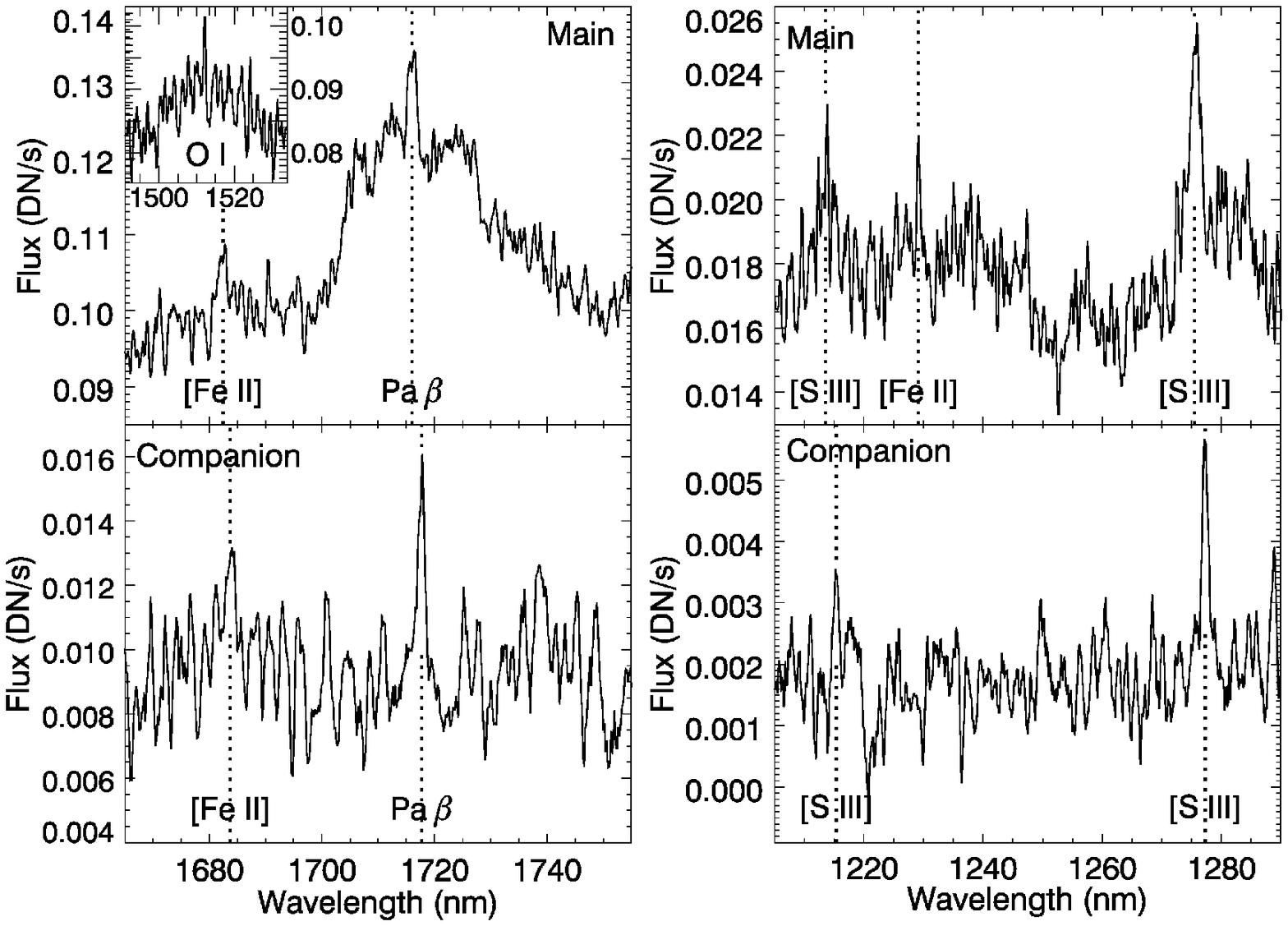}
\caption{ \textsl{H}- and \textsl{J}-band (left and right, respectively) spectra of the main galaxy (top) and the companion (bottom). The dotted lines mark the measured centers of the identified narrow lines.  The observed wavelengths are in units of vacuum $\mu$m.  We observe narrow emission lines (\siiiw, \feiij, \siiis, \oi, \feiih, \pab) as well as broad \pab~and broad \oi~from the main galaxy, making it a Type 1 AGN.  The broad \pab~has a measured FWHM of $6300\pm200$ \kms, and the broad \oi~has a measured FWHM of $5000\pm500$ \kms.  Only narrow emission lines (\siiiw, \siiis, \feiih, \pab) are evident in the companion, consistent with a Type 2 AGN.
}
\end{figure*}

We used a modified\footnote{Correcting for a $\sim6$\AA~shift of the previous wavelength calibration \citep{wright2011}.} version of the OSIRIS Data Reduction Pipeline (ODRP) v2.3\footnote{http://irlab.astro.ucla.edu/osiris/pipeline.html} to process our images.   Instead of using the standard mosaicking tools provided in the ODRP, we determined relative offsets between the images by fitting two-dimensional Lorentzian profiles to the main galaxy in each frame, and then input the relative offsets directly into the mosaicking module of the ODRP\footnote{See http://irlab.astro.ucla.edu/osiriswiki/dokuphp?id=mosaic \_with\_a\_list\_of\_offsets~for a detailed explanation.}.  Finally, we extracted and combined spectra from the five spatial pixels with the highest signal-to-noise ratio (S/N) of both the main galaxy and the companion in both \textsl{J} and \textsl{H} bands, as shown in Figure \ref{spectra}. Additionally, an OSIRIS image can be created by summing up the flux over all of the wavelengths (Figure \ref{images}, right panel).

\begin{table*}[t]
\begin{center}
\begin{tabular}{|c|cccc|cccc|c|}
\hline 
Spectral Line$^6$  & $z_{Main}$ & $\delta z_{M}^7$ &  $z_{Comp}$ & $\delta z_{C}^7$ & $v_{M}^8$ & {\normalsize $\delta v_{M}^{7,8}$} & $v_{C}^8$ & $\delta v_{C}^{7,8}$ & {\small $F_{M}/F_{C}^9$ } \\
\hline
\hline
\siiiw                                           & 0.33797   & 0.0003   & 0.33982 & 0.00014 & -25 & 63 & 413 & 31 & 5.7\\
\feii~$\lambda 9188$            & 0.33786   & 0.00007 &        ---    &    ---       & -27 & 16 & ---   & --- & --- \\
\siiis                                            & 0.33792 & 0.00013  & 0.33972 & 0.00008 & -14 & 28 & 391 & 18 & 5.3 \\
\feii~$\lambda 1.2570 \mu$m    & 0.33842 & 0.00017  &        ---    &    ---       & 98 & 38 & ---   & --- & --- \\
\pab~$\lambda 1.2822 \mu$m  & 0.33840 & 0.00009 & 0.33968 & 0.00011  & 93 & 20 & 382 & 24 & 2.3 \\
\hline
Averaged redshifts                   & 0.33810  & 0.00007 & 0.33974  & 0.00007 & 25 & 14 & 395 & 14 & --- \\
\hline
\hline
{\small \oiii~blue component}      & 0.33798 & 0.00004 &      ---    &       ---      &      0 &  9  & ---   & --- & 1.5 \\
{\small \oiii~red component }      &   ---        &    ---       & 0.33986  &  0.00004 &   --- &  ---  & 422 & 9 & --- \\
\hline
\end{tabular} \endgroup
\end{center}
\caption{ {\normalsize Redshifts} {\footnotesize~ $^6$Vacuum wavelengths. $^7$$1\sigma$ \textsc{measurement} errors. $^8$Velocities in \kms, measured with respect to the main galaxy's \oiii~blue component. $^9$Line flux ratio: main galaxy to companion galaxy.}}
\end{table*}

\section{Analysis } 

Spatially resolved spectroscopy will help us answer two questions: (1) do the redshifts of resolved spatial structures seen by NIRC2 and OSIRIS match the redshifts of the double peaks of the SDSS \oiiis~emission lines? and (2) What types of objects are the main galaxy and the companion: Type 1 AGN, Type 2 AGN, or a starburst?

\subsection{ Redshifts: Matching Spatial Structures to Spectral Double Peaks}

By fitting Gaussians, we measured central wavelengths for \siiiw, \feiij, \siiis, \oi, \feiih, and \pab, each labeled in Figure \ref{spectra}.  Table 1 shows the redshifts calculated from these wavelength centers, as well as the redshifts calculated from the SDSS double \oiii~components; Table 1 also includes the velocity offsets of these lines with respect to the main galaxy's \oiii~blue component.  Monte Carlo simulations of Gaussian peaks with superimposed random errors, with the same standard deviations as the noise in our spectra, allowed measurement of the fitting error.  The companion's narrow lines are on average 2.5 times more narrow than the main galaxy's narrow lines; this is consistent with the companion having a fainter host galaxy \citep{nelson1996}.
 
As shown in Table 1, the main galaxy's redshift ($0.33810\pm0.00007$) matches the SDSS \oiii~blue component redshift ($0.33798\pm0.00004$), while the companion's redshift ($0.33974\pm0.00007$) matches the SDSS \oiii~red component redshift ($0.33986\pm0.00004$).  This clearly illustrates that the double spatial structure corresponds to the double-peaked emission lines.   

\begin{figure}[t]
\figurenum{3}
\label{sdsswindow}
\centering
\includegraphics[width=\columnwidth]{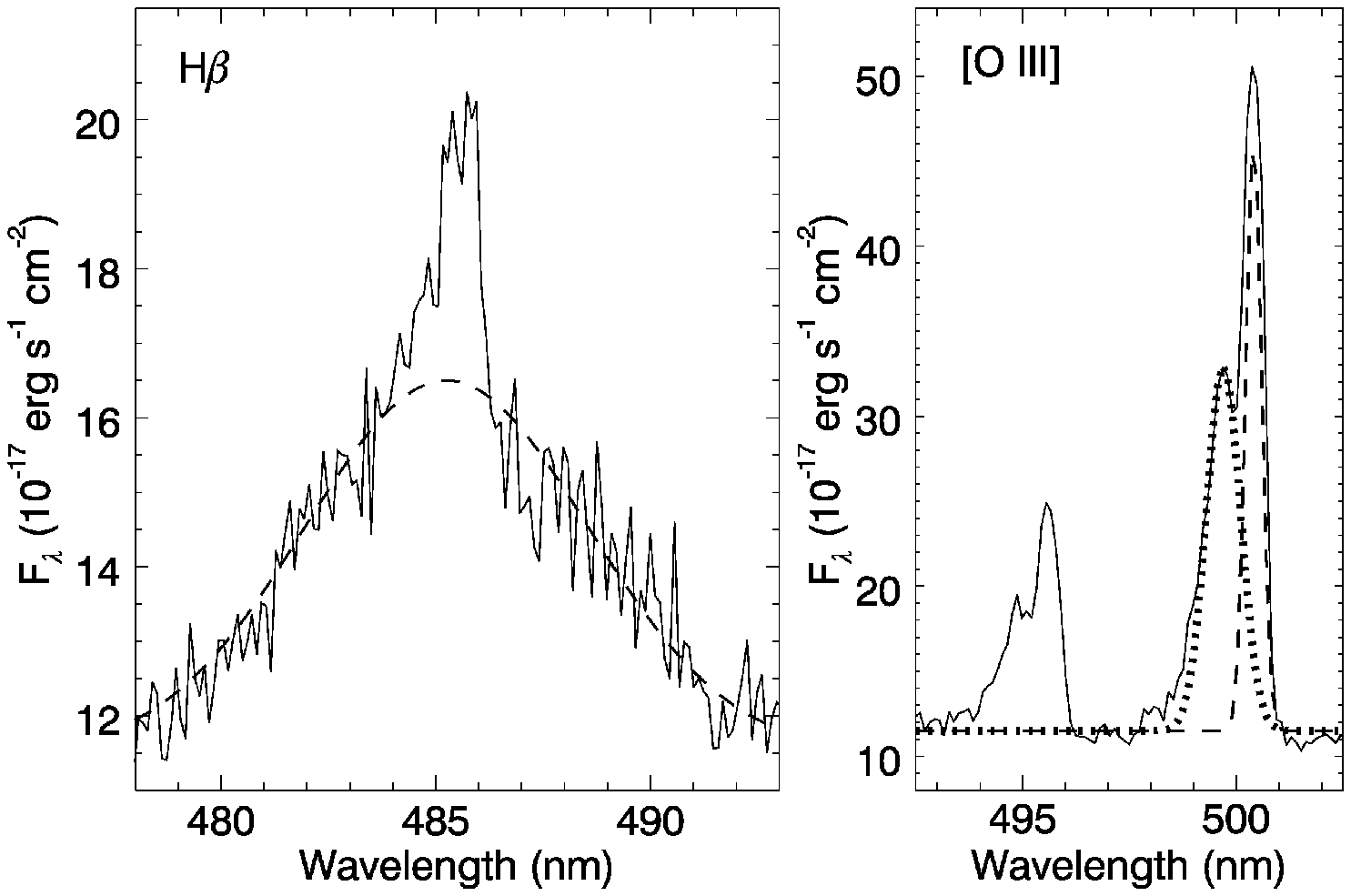}
\caption{ SDSS \obj~spectrum of (left) \hb~and (right) \oiiiw~and \oiiis~in rest vacuum wavelengths (\AA). In the left \hb~plot, the dashed curve is the fit to the broad component of the \hb~line; in the right \oiii~plot, the dotted and dashed lines are, respectively, the fits to the blue and red components of the narrow \oiiis~line.  The measured FWHM of \hb~is $5960\pm90$ \kms, which agrees well with the $6300\pm200$ \kms~FWHM measured from \pab~for the main galaxy.}
\end{figure}

However, the relative fluxes of the lines in the visible and in the NIR are not the same (Table 1).  For our IR narrow \siiiw~and \siiis~lines, the main galaxy is 5.3 and 5.7 times brighter than the companion, respectively, and in narrow \pab, the main galaxy is 2.3 times brighter than the companion.  In the visible SDSS \oiiis~line, the main galaxy's blue peak is 1.5 times stronger than the companion's red peak.  In Figure \ref{sdsswindow}, we show the SDSS spectrum of the \oiiiw, \oiiis, and \hb~lines; the overplotted fits show that although the \oiiis~red peak is taller, the \oiiis~blue peak is wider and thus contains more flux.  Since we matched the IR-bright main galaxy to the \oiii~blue peak and the IR-faint companion to the \oiii~red peak, there is clearly more going on with this system to create such varied flux ratios.   We suggest that the difference in visible and NIR flux ratios might be due to dust obscuring the main galaxy and not the companion, which would lead to lower visible flux ratios and higher NIR flux ratios in the main galaxy.

\subsection{ Identity of the Two Objects }

The presence of double peaks in spectral lines or of spatially-separated companion galaxies could be explained by a range of phenomena: chance superposition of two objects, a recoiling SMBH, jets, outflows from a single AGN, gravitationally lensed sources, starbursts, or rings of star formation.  Spatially resolved spectroscopy helps rule out many of these options.  Matching the individual components of the double-peaked emission lines to the two distinct galaxies shows that the double components do not come from rings.  As calculated in \citet{decarli2010}, the probability of a chance superposition for two AGNs with separations \textless 1\farcs5 with $0.35 < z < 0.45$ is $\sim2\times10^{-7}$, which is negligible with respect to the number of SDSS DR7 AGNs in this redshift range ($\sim 3300$).  Quasar clustering could give several orders of magnitude higher probability of a superposition, but a cluster is not evident surrounding \obj~in the SDSS image.  Even without considering the very close redshifts of the galaxies, the probability is so small that we can dismiss the possibility that the double lines or multiple imaged galaxies are due to a chance superposition.  Since \obj~is radio-quiet, these features are unlikely due to jets interacting with the surrounding medium.  Since the two galaxies have different spectra, the two galaxies are not two images of one gravitationally lensed object \citep[and references therein]{hennawi2006}.  

To distinguish between starbursts and AGNs of Types 1 and 2, we examine the observable broad lines and use several emission line ratio diagnostics.  In Figure \ref{spectra}, the spectrum for the main galaxy (top panels) clearly shows both a broad line for \pab~(left) and narrow emission lines (right).  The FWHM of this broad \pab~is $6300\pm200$ \kms, which roughly agrees with the $5960\pm90$ \kms~FWHM measured for the broad \hb~line in the SDSS spectrum (shown in Figure \ref{sdsswindow}).  We also see the broad \oi~$\lambda 1.12900 \mu m$ fluorescent line in our main galaxy's \textsl{H}-band spectrum.  The observed broad \oi, with its FWHM of $5000\pm500$ \kms, is observed in 67\% of Type 1 AGNs \citep{riffel2006}.  We measure the central wavelength of broad \oi~to be $1.5116\pm2 \mu m$, corresponding to a redshift of $0.3389\pm0.0018$; this redshift falls between with the average redshifts of the main galaxy and of the companion.   The presence of both broad and narrow lines indicates that the main galaxy is a Type 1 AGN.

\begin{figure}[t]
\figurenum{4}
\label{modBPT}
\centering
\includegraphics[width=\columnwidth]{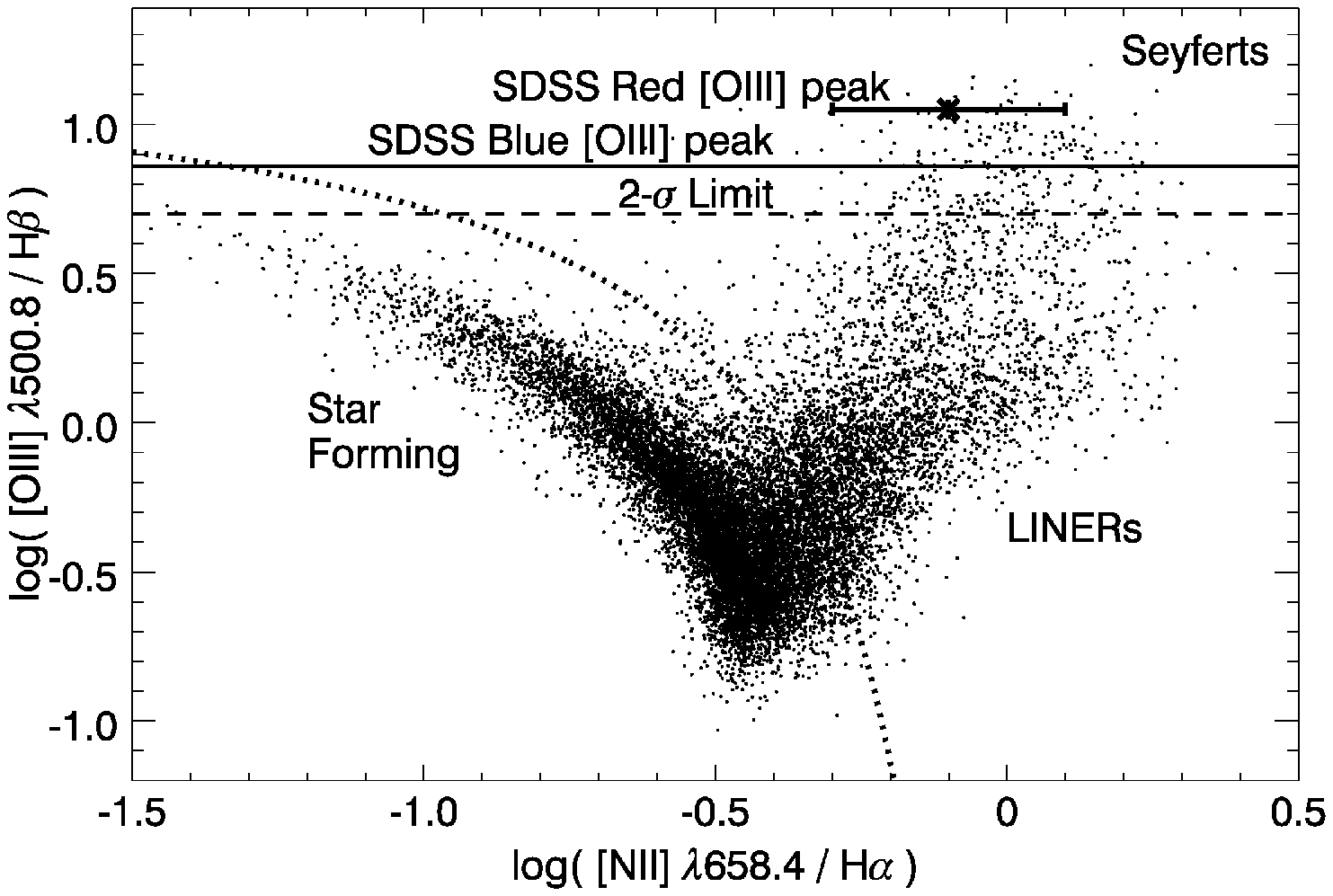}
\caption{ Baldwin--Philips--Terlevich diagram in which we plot the emission line flux ratios \oiiis/\hb~vs. the ratios \nii~$\lambda 6584$\AA/\ha~for a sample of 14,708 low redshift ($z$\textless0.3) SDSS galaxies in which all four lines are detected with S/N \textgreater~3.  The dotted curve shows the dividing line between starburst galaxies and AGNs as defined by \citet{kauffmann2003}.  The companion red-component line ratios are plotted by the asterisk symbol, with 1$\sigma$ error bars in the \nii/\ha~direction.  Since we are unable to disentangle an accurate measurement of the SDSS blue-component \nii/\ha, we plot the measured \oiii/\hb~flux ratio for the main galaxy's blue (solid) \oiii~component, as well as the 2$\sigma$ lower \oiii/\hb~limit of both the red and blue component ratios (dashed), as lines across the diagram.  Both \oiii~components fall above the location of star forming galaxies, meaning that both \oiii~components, and thus both the main galaxy and the companion, are due to AGNs.
}
\end{figure}

The companion shows only narrow lines, so it is not a Type 1 AGN.  The traditional way to distinguish between Type 2 AGNs and star-forming galaxies is to use measurements of \nii~$\lambda 6584$\AA/\ha~and \oiiis/\hb~to place the object on the Baldwin-Philips-Terlevich (BPT) diagram \citep[ and references therein]{bpt1981, kauffmann2003}.  Since the red component of the SDSS lines has a larger peak height, we were able to measure both \nii/\ha~and \oiii/\hb~for the companion.  Unfortunately, due to the complicated and noisy overlapping structure of the broad \ha~and double-peaked narrow \ha~and two \nii~lines, we are unable disentangle an accurate measurement of \nii/\ha~for the main galaxy.  However, the measurement of \oiii/\hb~can still be used to differentiate between a starburst companion and a Type 2 AGN.  As shown in Figure \ref{modBPT},\footnote[10]{SDSS galaxy line measurements from http://www.mpa-garching.mpg.de/ SDSS/DR7/raw\_data.html} both objects fall in the Seyfert/AGN region of the plot. For the main galaxy, we can combine the presence of broad lines in its spectrum with the fact that few star-forming galaxies exist at such high \oiii/\hb~ratios to conclude that it is a Type 1 AGN.  The line ratios of the companion place it clearly among the Seyferts, allowing us to conclude that the companion is a Type 2 AGN.

Using the OSIRIS spectra, we also use NIR emission line ratios for AGN and star formation diagnostics.  One such diagnostic pair is \feiih/\pab. \citet{rodriguezardila2008} showed that starbursts have $\log($\feii/\pab$) < -0.22$,  AGNs have $-0.22 < \log($\feii/\pab$) < 0.30$, and LINERs have $\log($\feii/\pab$) > 0.30$.  The main galaxy and the companion have $\log($\feii/\pab$)$ ratios, respectively, of $-0.07$ and $-0.16$; this provides further support that both the main galaxy and the companion are AGNs.

Another, weaker, indication that both galaxies are AGNs is that the spectra of both show narrow \siiiw~and \siiis~lines.  According to the AGN spectral atlas compiled by \citet{riffel2006}, all AGNs show \siii.  However, because the \siii~lines are also frequently seen in starburst galaxies, the presence of these lines does not clearly distinguish between AGNs and starbursts.

Our GALFIT analysis (Section 2) showed that both galaxies were well-fit by disk-like stellar light distributions for the host galaxies. Since both galaxies have disk morphologies, this suggests that both galaxies brought gas to the encounter. This is consistent with the fueling of both AGNs without requiring gas transfer between galaxies.

\section{Conclusions and Future work}

To confirm candidate dual AGNs selected via double-peaked \oiii~emission lines or multiple bright cores imaged in a single galaxy, spatially resolved spectroscopy is required.   For candidate dual AGNs with larger separations between components (\textgreater 1\farcs5), long slit spectrographs such as the Low Resolution Imaging Spectrometer can be used. For candidates with small separations (\textless1\farcs5), AO systems are best used to obtain spatially resolved spectra which clearly separated the contributions from the two galaxies.  We used the Keck 2 LGS AO system and the OSIRIS near-infrared integral field spectrograph to obtain spatially resolved spectra for \obj, a radio-quiet system with double-peaked \oiii~lines and two galaxies observed using Keck 2's LGS AO system and NIRC2.  The infrared redshifts of the two AGNs, separated by 4.8 kpc or 1\farcs0, match well with the redshifts of the SDSS \oiii~double peaks.  This directly links the double-peaked emission lines to the narrow-line regions of the two observed AGNs.  Line diagnostics indicate that both the companion and the main galaxy are bright due to AGN photoionization rather than through star formation or supernovae.  

Spatially resolved spectroscopic observations of further dual AGNs candidates are needed to construct a statistically-significant sample of true dual AGNs.  With such a sample, topics such as the merger rate or the AGN duty cycle can be addressed through comparisons with simulations of multiple mergers such as \citet{yu2011arxiv,lotza2010,lotzb2010}, and \citet{lotz2011arxiv}.

\acknowledgments
\textbf{Acknowledgments}

Data presented herein were obtained at the W. M. Keck Observatory,
which is operated as a scientific partnership among the
California Institute of Technology, the University of California,
and the National Aeronautics and Space Administration. The
Observatory and the Keck II Laser Guide Star AO system were both 
made possible by the generous financial support of the W. M. Keck 
Foundation. The authors wish to extend special thanks to those of
Hawaiian ancestry, on whose sacred mountain we are privileged
to be guests. Without their generous hospitality, the observations 
would not have been possible.

This material is based in part upon work supported by the National Science Foundation under award number AST-0908796. McGurk is supported by a Graduate Research Fellowship from the National Science Foundation.

Funding for the SDSS and SDSS-II has been provided by the Alfred P. Sloan Foundation, the Participating Institutions, the National Science Foundation, the U.S. Department of Energy, the National Aeronautics and Space Administration, the Japanese Monbukagakusho, the Max Planck Society, and the Higher Education Funding Council for England. The SDSS Web Site is http://www.sdss.org/.
  

\end{document}